\documentclass[english]{article}

\usepackage[T1]{fontenc}
\usepackage[latin9]{inputenc}
\usepackage{amssymb}
\usepackage{babel}

\usepackage[pdftex]{color,graphicx}

\begin{document}

\title{Late-time tails in extremal Reissner-Nordstrom spacetime}

\author{Amos Ori\footnote{\tt amos@physics.technion.ac.il}
\\
\\ 
\small{Department of Physics} \\ \small{Technion-Israel Institute of Technology} 
\\ \small{Haifa 32000, Israel}}

\date{\today}

  \maketitle

\begin{abstract}
This note discusses the late-time decay of perturbations outside extremal
Reissner-Nordstrom black hole. We consider individual spherical-harmonic
modes $l$ of massless scalar field. The initial data are assumed
to be of compact support, with generic regular behavior across the
horizon. The scalar perturbations are found to decay at late time
as $t^{-(2l+2)}$. We also provide the spatial dependence of the late-time
tails, including the exact overall pre-factor.
\end{abstract}

In this short note I describe the main results of a recent analysis
of late-time tails on extremal Reissner-Nordstrom (RN) spacetime,
described by the line element \[
ds^{2}=-(1-M/r)^{2}dt^{2}+(1-M/r)^{-2}dr^{2}+r^{2}d\Omega^{2}.\]
I consider a massless test scalar field $\Phi$, satisfying the standard
wave equation \begin{equation}
\square\Phi\equiv\Phi_{;\alpha}^{;\alpha}=0.\label{eq:Basic_Field}\end{equation}
In the analogous Schwarzschild case, four decades ago Price \cite{Price}
found a decay rate $t^{-2l-3}$ at late time, where $l$ is the mode's
multipolar number. The same decay rate was obtained soon afterward
by Bicak \cite{Bicak} for non-extremal RN. However, in the case of
extremal RN Bicak \cite{Bicak} obtained a much slower decay rate
$\propto t^{-l-2}$. 

The results presented here indicate a different decay rate in the
extremal case, $\propto t^{-2l-2}$, for generic initial data of compact
support. This applies (generically) as long as the initial support
intersects the horizon in a perfectly regular manner. (Otherwise,
if the initial data are not smooth enough at the horizon, the inverse-power
index may be smaller than $2l+2$. Alternatively, if the initial support
does not reach the horizon, the late-time decay will proceed in the
standard rate $\propto t^{-2l-3}$.) 

For a given multipole $l$ we define $\psi_{l}(r,t)$ as usual by
\[
\Phi=r^{-1}\psi_{l}(r,t)Y_{lm}(\theta,\varphi),\]
where $Y_{lm}$ denote the spherical harmonics. The field equation
(\ref{eq:Basic_Field}) then reduces to $\psi_{l}''-\ddot{\psi}_{l}=V_{l}(r)\psi_{l}$,
with the effective potential \begin{equation}
V_{l}(r)=\left(1-\frac{M}{r}\right)^{2}\left[\frac{2M}{r^{3}}\left(1-\frac{M}{r}\right)+\frac{l(l+1)}{r^{2}}\right].\label{eq:potential}\end{equation}
Here the prime and over-dot correspondingly denote differentiation
with respect to $r_{*}$ and $t$, where $r_{*}=r_{*}(r)$ is the
usual tortoise coordinate (which we set to vanish at $r=2M$). 

We introduce double-null coordinates ($u,v$) defined by $t=v+u$,
$r_{*}=v-u$. We use the characteristic initial-value formulation,
wherein the initial value of $\psi_{l}$ is specified along two intersecting
radial null rays. For convenience (and with no essential loss of generality)
we choose the two initial rays to be $u=0$ and $v=0$. We consider
initial data of compact-support. These initial data are further assumed
to be perfectly regular across the horizon. This implies that, when
expressed in terms of the variable $r$ (evaluated along the ingoing
ray $v=0$), the initial function $\psi_{u}^{l}(u)\equiv\psi_{l}(u;v=0)$
admits a regular Taylor expansion in the neighborhood of the horizon-crossing
point $r=M$: \begin{equation}
\psi_{u}^{l}(u)=c_{0}+c_{1}(r/M-1)+c_{2}(r/M-1)^{2}+...\label{eq:Horizon_Expansion}\end{equation}
where $c_{0},c_{1},c_{2}...$ are arbitrary expansion coefficients.
Our goal is to analyze the late-time behavior of $\psi_{l}$ which
evolves from such generic regular initial data. 

Our analysis is based on two essentially independent methods, both
leading to the same result. The first method takes advantage of the
conformal symmetry of the extremal RN spacetime \cite{conformal}
under the \emph{inversion transformation} $r_{*}\rightarrow-r_{*}$.
This transformation maps the horizon to FNI and vice versa. The effective
potential $V_{l}(r)$ is\emph{ invariant }under this inversion. Therefore,
given any solution $\psi_{l}(r,t)$ of the field equation, the inversion
transformation produces a new solution $\Psi_{l}(r_{*},t)\equiv\psi_{l}(-r_{*},t)$.
(An approximate version of this exact transformation was previously
used by Bicak \cite{Bicak} and later by Burko \cite{Burko}.) If
the initial support of $\psi_{l}$ is restricted to a small neighborhood
of the horizon, then the inversion transformation maps it to weak-field
initial data (at $r\gg M$) for $\Psi_{l}$. One can then apply the
approximate methods introduced by Price \cite{Price} (and more specifically,
the iterative scheme developed later by Barack \cite{Barack1}) to
analyze the evolving late-time tail of $\Psi_{l}$. This part of the
analysis was carried out by Or Sela and myself, and will be described
elsewhere \cite{long}. Then one applies the inversion transformation
again, obtaining the late-time behavior of the original field $\psi_{l}$.
One obtains, in the near-horizon domain ($-r_{*}\gg M$) 
\begin{equation}
\psi_{l}\cong e[M(1/u-1/v)]^{l+1},\label{eq:psi_HOR}\end{equation}
(presumably with additional terms of higher order in $1/u$ and/or
$1/v$), where $e$ is a certain coefficient {[}determined by the
initial parameters in Eq. (\ref{eq:Horizon_Expansion}){]}. 

In the second method we exploit the exact conserved quantities recently
found by Aretakis \cite{Aretakis}. He showed that for each $l$,
a certain differential operator $d_{l}$ (applied to $\psi_{l}$)
is conserved along the horizon of extremal RN. Assuming a rather generic
combination of inverse powers of $u$ and $v$ near the horizon, and
applying $d_{l}$ to this combination {[}and also to the initial conditions
(\ref{eq:Horizon_Expansion}){]}, one arrives at the expression (\ref{eq:psi_HOR}),
along with an exact expression for the pre-factor: \begin{equation}
e=2\,\frac{[(l+1)!]^{2}}{(2l+2)!}\,(c_{l+1}+\frac{l}{l+1}c_{l}).\label{eq:Exact_Coefficient}\end{equation}

In the final stage we transform the inverse powers of $u$ and $v$
to inverse powers of $t.$ To this end we employ the late-time expansion,
presented (for somewhat different situations) in e.g. Refs. \cite{Late-time}
and \cite{Barack2}. The final result is 
\begin{equation}
\Psi_{l}=(-4)^{l+1}\, e\, M^{3l+2}\,[r(r-M)^{-l-1}]\, t^{-2l-2}\:+\: O(t^{-2l-3}),\label{eq:Large_t}\end{equation}
applicable throughout the large-$t$ domain $t\gg|r_{*}|$.

We point out that this decay rate applies when the initial support
intersects the horizon. If, on the other hand, the initial support
is disjoint from the horizon (in which case $c_{l}=c_{l+1}=0$), then
the late-time decay will proceed in the usual rate $t^{-2l-3}$.

Burko \cite{Burko} considered the problem of tails on extremal RN
background for non-generic initial data of two specific types: (i)
compact initial support which does \emph{not} extend up to the horizon,
and (ii) initially-static multipoles. In both cases, his results are
fully consistent with Eqs. (\ref{eq:Exact_Coefficient},\ref{eq:Large_t})
above. 

Recently Lucietti et al. \cite{Lucietti} numerically integrated the
field equation (\ref{eq:Basic_Field}) on the extremal RN background,
and analyzed the late-time behavior for $l=0,1,2$. They obtained
the desired decay rates $\propto t^{-2l-2}$ in all three cases. It
will be interesting to numerically test the $r$-dependence as well
as the pre-factor in our final result (\ref{eq:Large_t})---and also
to extend the numerical investigation to additional $l$ values. 

I would like to thank Stefanos Aretakis and Or Sela for numerous interesting
and helpful discussions. I am especially grateful to Mihalis Dafermos
for generous hospitality during my visits in Princeton university
and in Cambridge university, where this research was initiated, and
for many interesting conversations during these visits.

\end{document}